\documentclass[aps,notitlepage,superscriptaddress]{revtex4-1}
\usepackage{graphicx}
\usepackage{amssymb,amsfonts,amsmath}
\usepackage[colorlinks=true,citecolor=Cerulean,linkcolor=RubineRed,urlcolor=Cerulean]{hyperref}
\hypersetup{breaklinks=true}
\usepackage{graphicx}
\usepackage{color}
\usepackage[usenames,dvipsnames]{xcolor}
\usepackage{epstopdf}
\usepackage{braket}
\usepackage{physics}
\usepackage[margin=2cm]{geometry}
\usepackage{dsfont}
\usepackage{float}
\usepackage{subfig}

\newcommand{\gfunction}{g(k_\ell, k_m, k_n, k_p)}
\newcommand{\gammazero}{\Gamma(k_\ell, k_m, k_n, k_p)}
\newcommand{\gfunctionPhi}{g(k_\ell, k_m, k_n, k_p;\Phi)}
\newcommand{\bessel}{\mathcal{J}_0\left[2\kappa\gfunction\right]}
\newcommand{\besselPhi}{\mathcal{J}_0\left[2\kappa\gfunctionPhi\right]}

\begin{document}
\title{Cat states in a driven superfluid: role of signal shape and switching protocol}

\author{Jes\'us Mateos}
\affiliation{Departamento de F\'isica de Materiales, Universidad
Complutense de Madrid, E-28040 Madrid, Spain}

\author{Gregor Pieplow}
\affiliation{Departamento de F\'isica de Materiales, Universidad
Complutense de Madrid, E-28040 Madrid, Spain}
\affiliation{Department of Physics, Humboldt-Universit\"at zu Berlin, Newtonstr. 15, D-12489 Berlin, Germany}

\author{Charles Creffield}
\affiliation{Departamento de F\'isica de Materiales, Universidad
Complutense de Madrid, E-28040 Madrid, Spain}

\author{Fernando Sols}\thanks{f.sols@ucm.es}
\affiliation{Departamento de F\'isica de Materiales, Universidad
Complutense de Madrid, E-28040 Madrid, Spain}

\date{\today}

\begin{abstract}
We investigate the behavior of a one-dimensional Bose-Hubbard model whose kinetic energy is made to oscillate with zero time-average. The effective dynamics is governed by an atypical many-body Hamiltonian where only even-order hopping processes are allowed. At a critical value of the driving, the system passes from a Mott insulator to a superfluid formed by a cat-like superposition of two quasi-condensates with opposite non-zero momenta. We analyze the robustness of this unconventional ground state against variations of a number of system parameters. In particular we study the effect of the waveform and the switching protocol of the driving signal. Knowledge of the sensitivity of the system to these parameter variations allows us to gauge the robustness of the exotic physical behavior.
\end{abstract}

\maketitle
\section{Introduction}
\label{sec:1}
``Floquet engineering'' \cite{engineering}
consists of rapidly oscillating a parameter of a
Hamiltonian periodically in time, which, following the elimination
of the high-frequency degrees of freedom, produces a time-independent
effective Hamiltonian. The process can both produce new terms in
the effective Hamiltonian which
do not appear in the undriven model, or the renormalization of 
previously existing processes.
A well-known example of the latter is the periodically-shaken
Bose-Hubbard model \cite{holthaus,creff}, 
in which the lattice-shaking causes a renormalization of 
the single-particle tunneling.
In principle, any term in the
Hamiltonian can be periodically varied, and Floquet theory used to derive
the resulting effective model. The most common forms of driving are the
variation of an external potential -- which includes, for example, the case
of lattice-shaking mentioned previously -- and, more recently,
the oscillation of the interparticle interactions \cite{santos,cristiane}.

In Ref. \cite{kinetic_NJP} we introduced a new form of driving which we
term ``kinetic driving''. This involves the periodic modulation of 
the kinetic component of a system's Hamiltonian, or equivalently, 
varying the system's hopping parameter. Small modulations of the
hopping in the Hubbard model have already been used as a tool
to probe the system's properties \cite{esslinger,kollath,demler,demler_recent}. In contrast, we
considered the hopping parameter to have the time-dependent form
$J(t) = J \cos \omega t$, so that its time-averaged value {\em vanishes}. 
As a consequence, particles can only move via higher-order
processes, such as the hopping of particles in pairs (doublons), or
by long-range single-particle hopping processes assisted by the presence of another particle elsewhere.
As a result 
instead of condensing at zero momentum, like a standard
Bose-Einstein condensate, the system condenses into a
macroscopic superposition of two condensates with
non-zero momenta, $\pm \pi/2$. In Ref. \cite{kinetic_PRR} we showed
that this superposition has the form of a Schr\"odinger cat state,
and the unusual properties of the system give it remarkable stability.

In this work we extend our study of the properties of this exotic
superfluid state. After a brief description of preceding work in Section \ref{sec:2}, we consider in Section \ref{sec:3} non-sinusoidal driving functions and study how the shape of the waveform affects the response of
the system. We then go on to include a temporal phase-shift in
the driving in Section \ref{sec:4}, to further characterize the stability of the cat state to
perturbations of the driving potential, and in Section \ref{sec:5} we study the response
of the system to an external flux. Finally in Section \ref{sec:6} we consider the
preparation of the cat state in experiment by adiabatically
ramping the driving from zero, thereby evolving the system from
a Mott state to the superfluid state, and show how the ramp time  
depends on the choice of the driving function.

\section{Model}
\label{sec:2}

We consider the time-dependent Bose-Hubbard Hamiltonian

\begin{equation}
\mathcal{H}(t) = -J f(t) \sum_{x=0}^{L-1} 
\left( a_x^\dagger a_{x+1} + a_{x+1}^\dagger a_x \right)
+ \frac{U}{2} \sum_{x=0}^{L-1} n_x(n_x-1)~,
\label{eq:driven_BH}
\end{equation}
where $a_x$($a_x^\dagger$) are the usual bosonic annihilation(creation) operators and $n_x = a_x^\dagger a_x$ is the number operator.
The Hubbard interaction energy is given by $U>0$, and $f(t)$ is the 
$T$-periodic function modulating the hopping amplitude $J$ between nearest-neighbor sites.

We may introduce the plane-wave representation
\begin{align}
\label{eq:planewaves}
a_x=\frac1{\sqrt{L}}\sum_{\ell=0}^{L-1}e^{ik_\ell x}a_{k_\ell},\quad
a_{k_\ell}=\frac1{\sqrt{L}}\sum_{x=0}^{L-1}e^{-ik_\ell x}a_x,
\end{align}
where $k_\ell=2\pi\ell/L$. 
In Refs. \cite{kinetic_NJP,kinetic_PRR} it was proven that for high frequencies the Hamiltonian \eqref{eq:driven_BH} can be 
accurately approximated by an effective time-independent Hamiltonian, obtained by making a unitary transformation to the interaction picture, and averaging over one period of the driving
\begin{align}
\label{eq:Heffgen}
H_\mathrm{eff}
=\frac{U}{2L}\sum_{\ell,m,n,p=0}^{L-1}\delta_{k_\ell+k_m,k_n+k_p}\Gamma(k_\ell,k_m,k_n,k_p)a^\dagger_{k_p}a^\dagger_{k_n}a_{k_m}a_{k_\ell}\, ,
\end{align}
where
\begin{align}
\label{eq:gamma0}
\gammazero=\frac1{T}\int_0^Tdt e^{2iF(t)\kappa \gfunction} \equiv \Gamma(\kappa g),
\end{align}
\begin{align}
\label{eq:g}
g(k_\ell, k_m, k_n, k_p) \equiv \cos (k_\ell)+\cos (k_m)-\cos (k_n)-\cos (k_p) .
\end{align}
In Eq.\eqref{eq:gamma0} we introduce the primitive of the driving potential
$F(t)=\omega\int_0^t f(t')dt'$, and in this work we parameterize the driving using the dimensionless quantity $\kappa=J/\omega$, where $\omega = 2\pi/T$. For cosenoidal driving,
$f(t)=\cos(\omega t)$, the Hamiltonian \eqref{eq:Heffgen} becomes \cite{kinetic_PRR}
\begin{align}
\label{eq:Heffcos}
H_\mathrm{eff}
=\frac{U}{2L}\sum_{\ell,m,n,p=0}^{L-1}\delta_{k_\ell+k_m,k_n+k_p}\bessel a^\dagger_{k_p}a^\dagger_{k_n}a_{k_m}a_{k_\ell},
\end{align}
where $\mathcal{J}_0$ is the zeroth-order Bessel function of the first kind.

We note that in this derivation we require $\kappa < 1$, to be consistent with the high-frequency approximation.

\section{Signal shape}
\label{sec:3}

\begin{figure}[t]
\resizebox{1.0\columnwidth}{!}{%
\subfloat{\includegraphics[scale=0.8]{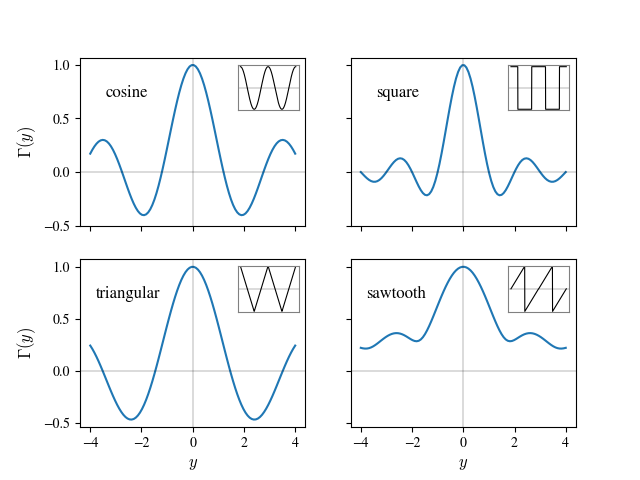}}}
\caption{The matrix element defining the effective interaction between bosons, $\Gamma$, for the various signal shapes, plotted as a function of the variable $y=\kappa\gfunction$. The analytical expressions for $\Gamma(y)$ are given in Section \ref{sec:3}. For the sawtooth case, where $\Gamma$ cannot be made real (see discussion in Sections \ref{sec:3} and \ref{sec:4}), we plot $|\Gamma(y)|$.}
\label{fig:shapes}       
\end{figure}

In this Section we will derive the expression of the effective Hamiltonian \eqref{eq:Heffgen} for driving functions $f(t)$ other than the cosenoidal signal, but still periodic and with zero time-average over one period. These signals are shown versus time in the insets of Fig. \ref{fig:shapes}. They are taken to have the same period and amplitude.

Equation \eqref{eq:Heffgen} remains the general effective Hamiltonian, but different profiles of the time signal $f(t)$ now yield different matrix elements of the interaction between plane waves, i.e., different forms of the function $\Gamma(y)$, where $y=\kappa\gfunction$. We have already seen that for the cosenoidal signal [see Eq. \eqref{eq:Heffcos}], $\Gamma(y)=\mathcal{J}_0 (2y)$. 

For the square wave signal
\begin{align}
\label{eq:squaref}
f(t)=\left\{\begin{array}{ll}
1, & \omega t\in[0,\pi/2)\\
-1, & \omega t\in[\pi/2,3\pi/2)\\
1, & \omega t\in[3\pi/2,2\pi)\\
\end{array}
\right.,
\end{align}
the matrix element $\Gamma(y)$ \eqref{eq:gamma0} takes the form
\begin{align}
\label{eq:squaregamma}
\Gamma(y)= \frac{\sin (\pi y)}{\pi y} \, ,
\end{align}
while for  the triangular signal
\begin{align}
\label{eq:trianf}
f(t)=\frac{2}{\pi}\times\left\{\begin{array}{ll}
\pi/2-\omega t, & \omega t\in[0,\pi)\\ 
\omega t -3\pi/2, &\omega t\in[\pi,2\pi)\\
\end{array}
\right.,
\end{align}
the expression becomes
\begin{equation}
\label{eq:triangamma}
\Gamma(y) = \frac{\cos\left(\dfrac{\pi y}{2}\right) C(\sqrt{y}) 
+ \sin\left(\dfrac{\pi y}{2}\right) S(\sqrt{y}) } {\sqrt{y}} \, ,
\end{equation}
where $C$ and $S$ are the Fresnel cosine and sine integrals \cite{abramowitz}.

\begin{figure}[t]
\resizebox{0.9\columnwidth}{!}{%
\subfloat{\includegraphics[scale=0.8]{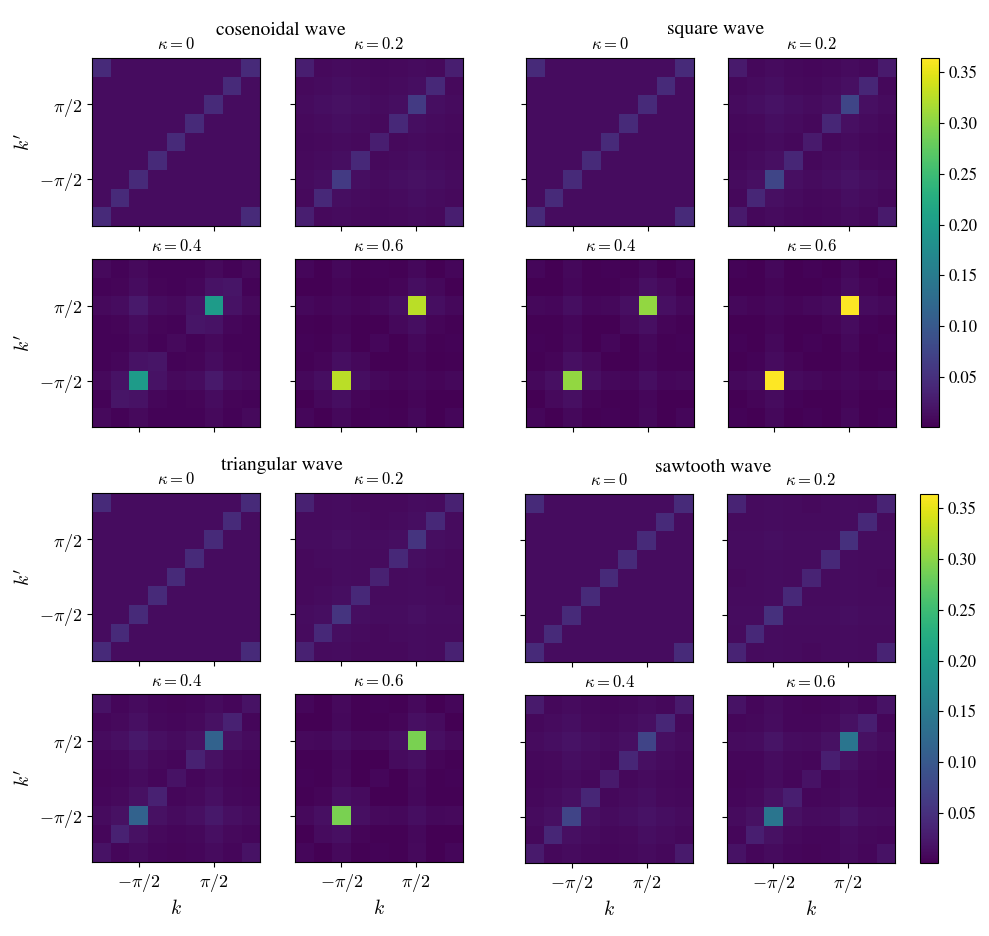}}}
\caption{Two particle momentum density, $\rho^{(2)}(k,k')$ as a function of $\kappa$ for each signal shape. For $\kappa=0$ the system is in the Mott state, and $\rho^{(2)}(k,k')$ is peaked along
the diagonal $k = k'$. As $\kappa$ increases, isolated peaks form at $\pm(\pi/2, \pi/2)$, 
indicating the formation of the superfluid cat-state.
} 
\label{fig:rho2_shapes}       
\end{figure}

Finally, for the sawtooth signal
\begin{align}
\label{eq:sawf}
f(t)=\frac{1}{\pi}\times\left\{\begin{array}{ll}
\omega t, & \omega t\in[0,\pi)\\ \\
\omega t-2\pi , & \omega t\in[\pi,2\pi)
\end{array}
\right.,
\end{align}
we obtain
\begin{align}
\label{eq:sawgamma}
\Gamma (y) = \frac{\mathrm{erf}\left( \sqrt{-i\pi y} \right)}
{2\sqrt{-i y}},
\end{align}
where $\text{erf}(z)$ is the error function. 

Figure \ref{fig:shapes} shows the plots of $\Gamma(y)$ for the different driving profiles. 
Since $\Gamma$ is complex for the sawtooth signal, we plot $|\Gamma|$ in that case.
For all signals we have $\Gamma(0)=1$. 

The non-vanishing of $\Gamma$ for the sawtooth signal can be directly ascribed to its lack of time-reversal symmetry. Unlike the other waveforms we consider, the sawtooth function is not symmetric under the generalized parity operation $(x,t) \rightarrow (-x,t+T/2)$. As a consequence, the quasienergies of 
$H_\mathrm{eff}$ do not have definite parity \cite{grossmann}, and so the von Neumann-Wigner theorem \cite{vonneumann} forbids them making exact crossings as $\kappa$ varies. Instead, they can only form avoided crossings, which in turn implies that the amplitudes for elementary processes, described by
$\Gamma(y)$, cannot vanish.  

From \eqref{eq:Heffgen}, and using the specific expression of $\Gamma(y)$ of each signal, we calculate the two-particle momentum density
\begin{align}
\rho^{(2)}(k,k')=\frac1{N^2}\braket{n_k n_{k'}}.
\label{eq:rho2}
\end{align}
Results are shown in Fig. \ref{fig:rho2_shapes}. All the plots reveal the system is robust against the choice of the driving profile, giving very similar forms for $\rho^{(2)}(k,k')$. The only difference is that for the sawtooth, greater values of $\kappa$ are needed to produce peaks with comparable heights
to the other cases, which is related to the fact that $\Gamma(y)$ for the sawtooth does not cross the horizontal axis (see Fig. \ref{fig:shapes}).

\begin{figure}
\resizebox{1.0\columnwidth}{!}{%
\includegraphics{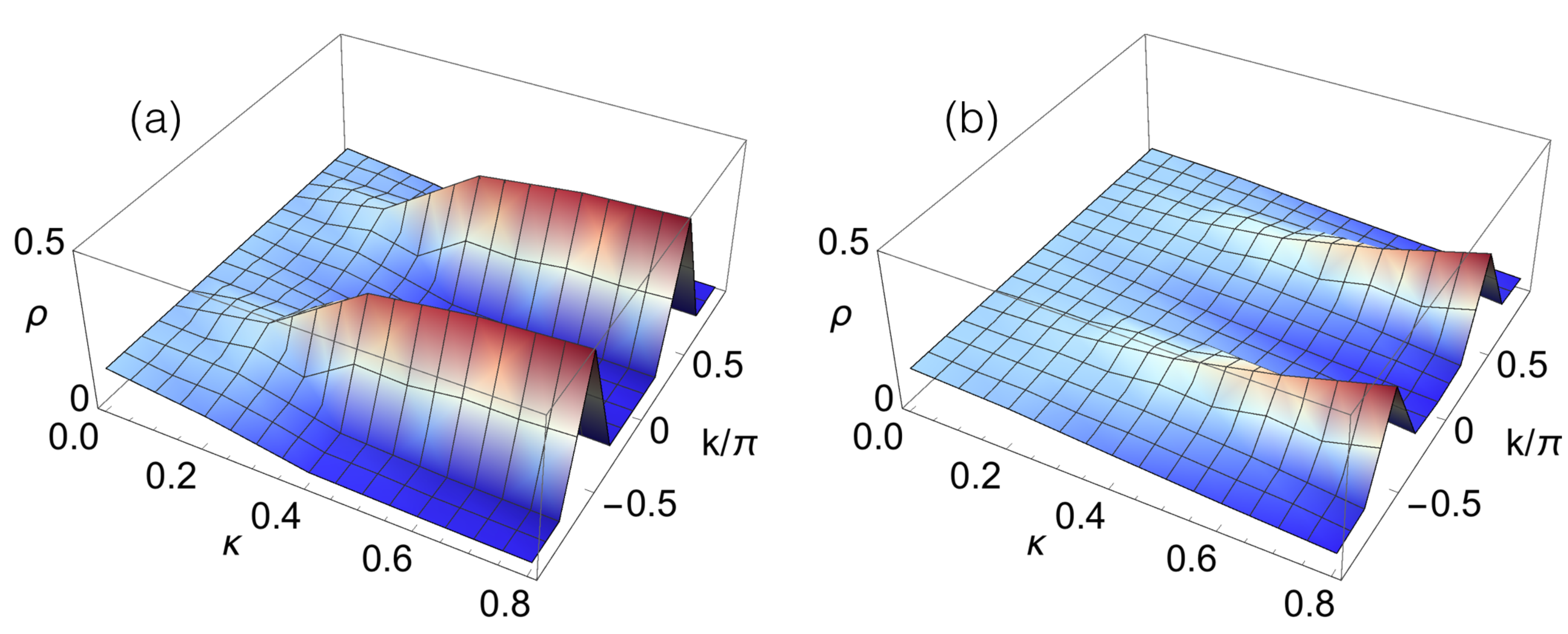}}
\caption{Momentum density, $\rho(k)$, as a function of $\kappa$.
(a) Square wave driving. The momentum density is initially
flat for $\kappa=0$ when the system is in the Mott state. As $\kappa$ 
increases the system passes through a phase
transition and $\rho(k)$ develops two peaks at $k = \pm \pi/2$, indicating
the formation of the superfluid cat state.
(b) Sawtooth driving. In contrast to the square wave case, the peaks
in $\rho(k)$ develop more slowly, and are much less pronounced.} 
\label{fig:mom_den}       
\end{figure}

We now calculate the one-particle momentum density
\begin{align}
\label{eq:rho}
\rho(k)=\frac1{N}\braket{n_k}.
\end{align}
for the square and the sawtooth signals, and show the results in Fig. \ref{fig:mom_den}. Again, peaks form at $k = \pm \pi/2$ when $\kappa$ grows, indicating the macroscopic occupation of these momentum states 
as the superfluid state develops. For the sawtooth these peaks in $\rho(k)$ develop more slowly as a function of $\kappa$, showing that larger values of $\kappa$ are required for the system to become superfluid.

In all the cases considered in this Section we have explicitly checked that the full time-dependent evolution under Eq. \eqref{eq:driven_BH} yields the same results as that obtained from the effective time-independent Hamiltonian $H_\mathrm{eff}$. 
We note that in the numerical simulation of the time evolution, the system is prepared in the Mott state (i.e. the ground state of the zero-hopping Bose-Hubbard Hamiltonian). The kinetic driving is then gradually introduced so that the state evolves adiabatically towards the ground state of the effective time-independent Hamiltonian.

\section{Initial phase}
\label{sec:4}

It was noted in Refs. \cite{creff_sols_PRA11,kudo,kolovsky,creff_sols_EL,creff_sols_PRA14},
and later observed in cold atoms systems \cite{haller},
that if a phase-shift $\varphi$ is added to the standard cosenoidal signal
\begin{equation}
f(t)=\cos(\omega t + \varphi) \, ,
\label{cos-standard-signal-phase}
\end{equation}
then the long-term dynamics can be sensitive to that phase and, in the case of a ring this phase can create the effect of an effective flux threading the ring, enabling the simulation of a synthetic magnetic field. We wish to investigate whether a similar effect appears in the kinetically-driven system. The effective time-independent Hamiltonian becomes \cite{goldman}
\begin{align}
H_\mathrm{eff}
=\frac{U}{2L}\sum_{\ell,m,n,p=0}^{L-1}\delta_{k_\ell+k_m,k_n+k_p}&e^{-i2 \kappa g F(\varphi)}
\Gamma(\kappa g)
\ a^\dagger_{k_p}a^\dagger_{k_n}a_{k_m}a_{k_\ell} \, ,
\label{eq:Heffphi}
\end{align}
where $\Gamma(y)$ is given in the previous Section for the various signal shapes, and the $F$ function reads
\begin{align}
F(\varphi) &= 
\begin{cases}
\sin \varphi \qquad \quad & (\mathrm{cosine \ wave})  \\
\varphi & (\mathrm{square \ wave})  \\
\varphi -\varphi^2/\pi & (\mathrm{triangular \ wave})  \\
\varphi^2/2\pi & (\mathrm{sawtooth \ wave}) \, .
\end{cases}
\end{align}

Due to the additive structure of the function $g$ [see Eq. \eqref{eq:g}], the presence of $F(\varphi) \neq 0$ amounts to adding a phase to each boson operator
\begin{equation}
\label{global-phase-k}
a_k \rightarrow a_k e^{i\alpha_k} \, ,
\end{equation}
where $\alpha_k =2 \kappa \cos(k) F(\varphi)$. Addition of a global, $k$-dependent phase to each one-particle momentum eigenstate of eigenvalue $k$ does not have any physical consequence, from which we conclude that the system properties are exactly independent of $\varphi$, as can be confirmed numerically.

Thus, unlike the single-particle hopping of independent particles, or in the conventional Bose-Hubbard problem \cite{creff_sols_PRA11,creff_sols_EL,creff_sols_PRA14}, it is not possible to create a flux through the ring by tailoring the initial phase of the kinetic driving.

\begin{figure}[t!]
\includegraphics[width = 1.1\textwidth]{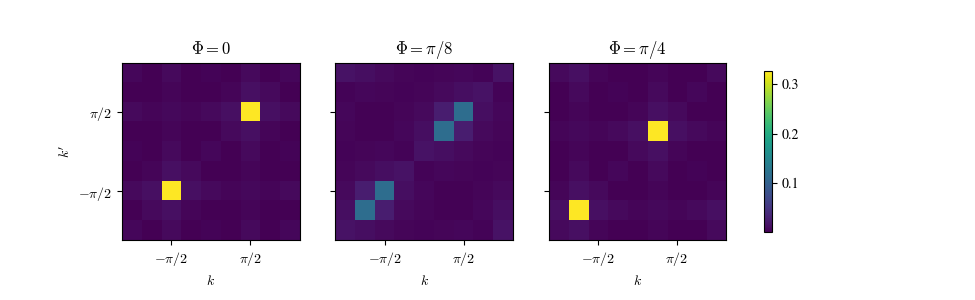}
\caption{Two-particle momentum density $\rho^{(2)}(k,k')$ for several values of $\Phi$, with $\kappa=0.6$. When $\Phi=0$ the peaks are centered on $\pm(\pi/2,\pi/2)$. For $\Phi = \pi/8$ each peak is smeared over two values of momenta, as the momentum shift produced by $\Phi$ is incommensurate with the reciprocal lattice momenta. $\Phi = \pi/4$ is commensurate with the reciprocal lattice and the peaks are shifted by one momentum spacing, $-\pi/4$, along the diagonal.}
\label{fig:Phi}
\end{figure}

\section{Effective flux}
\label{sec:5}

An effective external flux $\Phi$ threading the ring may result from a controlled or spurious rotation of the ring. We study here how ground state properties of the system change under the effect of this flux. The fundamental time-dependent Bose-Hubbard Hamiltonian becomes
\begin{align}
\label{eq:HgeneralPhi}
\mathcal{H}(t)=-f(t)\sum_{x=0}^{L-1}\left(e^{i\Phi}a^\dagger_{x+1}a_x+
e^{-i\Phi}a^\dagger_{x}a_{x+1}\right)+\frac{U}{2}\sum_{x=0}^{L-1}n_x(n_x-1),
\end{align} 
where we take the cosenoidal driving without any initial phase, $f(t)=\cos(\omega t)$. In this case the effective Hamiltonian becomes
\begin{align}
\label{eq:HeffPhi}
H_\mathrm{eff}(\Phi)
=\frac{U}{2L}\sum_{\ell,m,n,p=0}^{L-1}\delta_{k_\ell+k_m,k_n+k_p}\besselPhi a^\dagger_{k_p}a^\dagger_{k_n}a_{k_m}a_{k_\ell}
\end{align}
where the $g$-function is now defined by
\begin{align}
\label{eq:gPhi}
\gfunctionPhi=\cos (k_\ell+\Phi)+\cos (k_m+\Phi)-\cos (k_n+\Phi)-\cos (k_p+\Phi).
\end{align}
This is the main difference between this case and the previous one; $\Phi$ does not appear
as a phase factor, but inside the arguments of the cosine functions. Accordingly $\Phi$
does have a physically observable effect, shifting the momentum at which the
peaks of $\rho^{(2)}(k,k')$ occur.
In Fig. \ref{fig:Phi} (left) we shown the two-particle momentum density for $\Phi=0$, which
displays narrow peaks centered on $\pm(\pi/2, \pi/2)$. When a phase of $\Phi=\pi/8$ is introduced, Eq. \eqref{eq:gPhi} predicts these peaks should be shifted in momentum space by $\Delta k = \pi/8$.
However, for the 8-site ring we consider, this shift
is not commensurate with the reciprocal lattice momenta, and so the peaks spread over the two closest momenta to the shifted values, as shown in  Fig. \ref{fig:Phi} (center). 
The smallest non-zero shift of momentum that is commensurate with the lattice is 
$\Delta k=2\pi/8=\pi/4$. 
For $\Phi=\pi/4$, the peaks are indeed shifted by this quantity, so they now are located at $(-3\pi/4,-3\pi/4)$ and $(\pi/4, \pi/4)$, as can be seen in  Fig. \ref{fig:Phi} (right).

The fact that for $\Phi =\pi/8$ we find a preferential occupation of the available momenta which are closest to the ideal one reflects the physical continuity of the occupation as a function of momentum, which is somewhat obscured in this model with just a few momentum states. In the thermodynamic limit we can expect to find a smoothly enhanced occupation of momenta near $\pm \pi/2$, and a hint of this behavior can already be observed in Fig. \ref{fig:mom_den}, where an enhancement of the occupation is already observed in momenta near $\pm \pi/2$.

\begin{figure}
\resizebox{1.2\columnwidth}{!}{%
\includegraphics{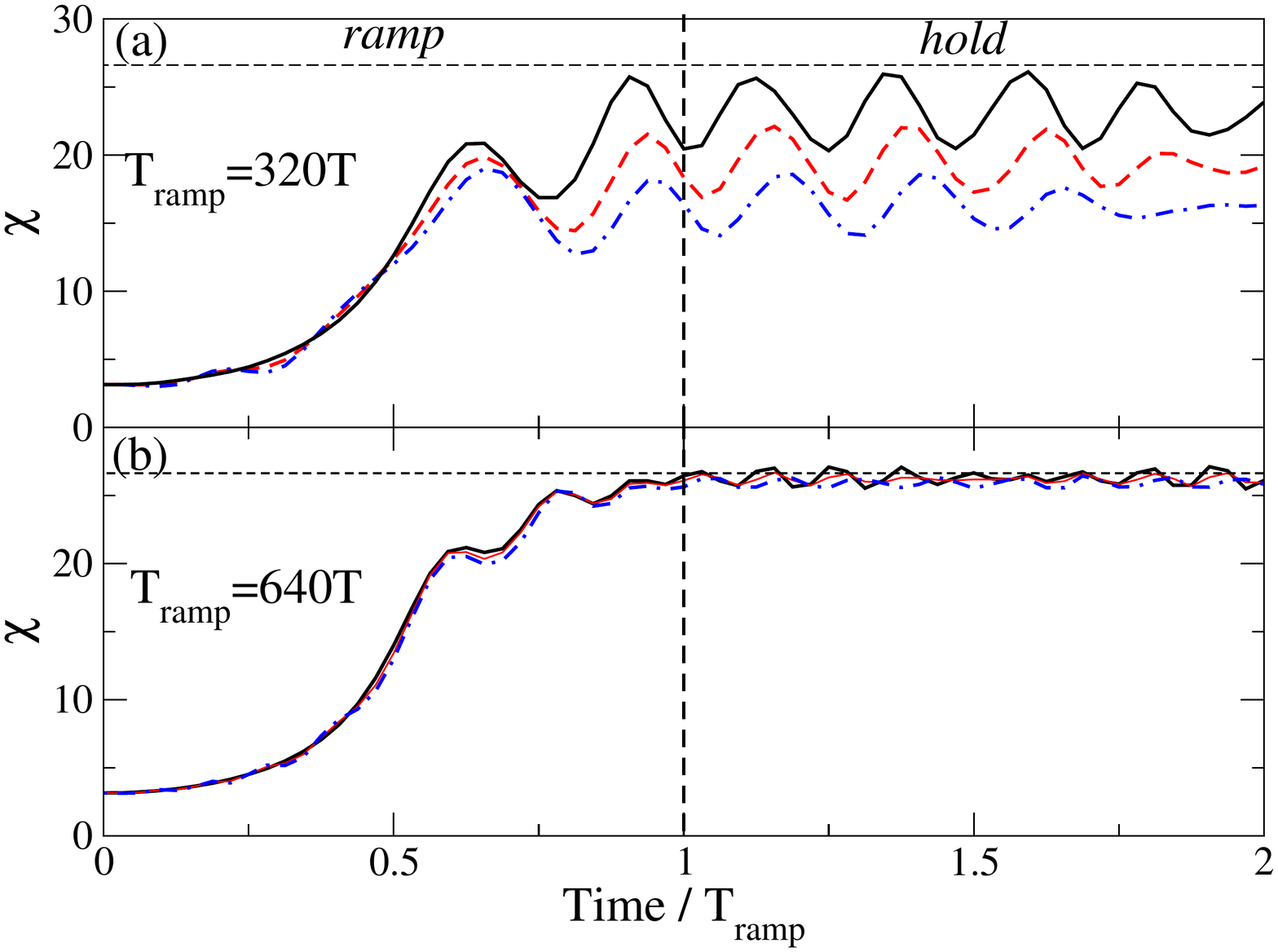} 
\includegraphics{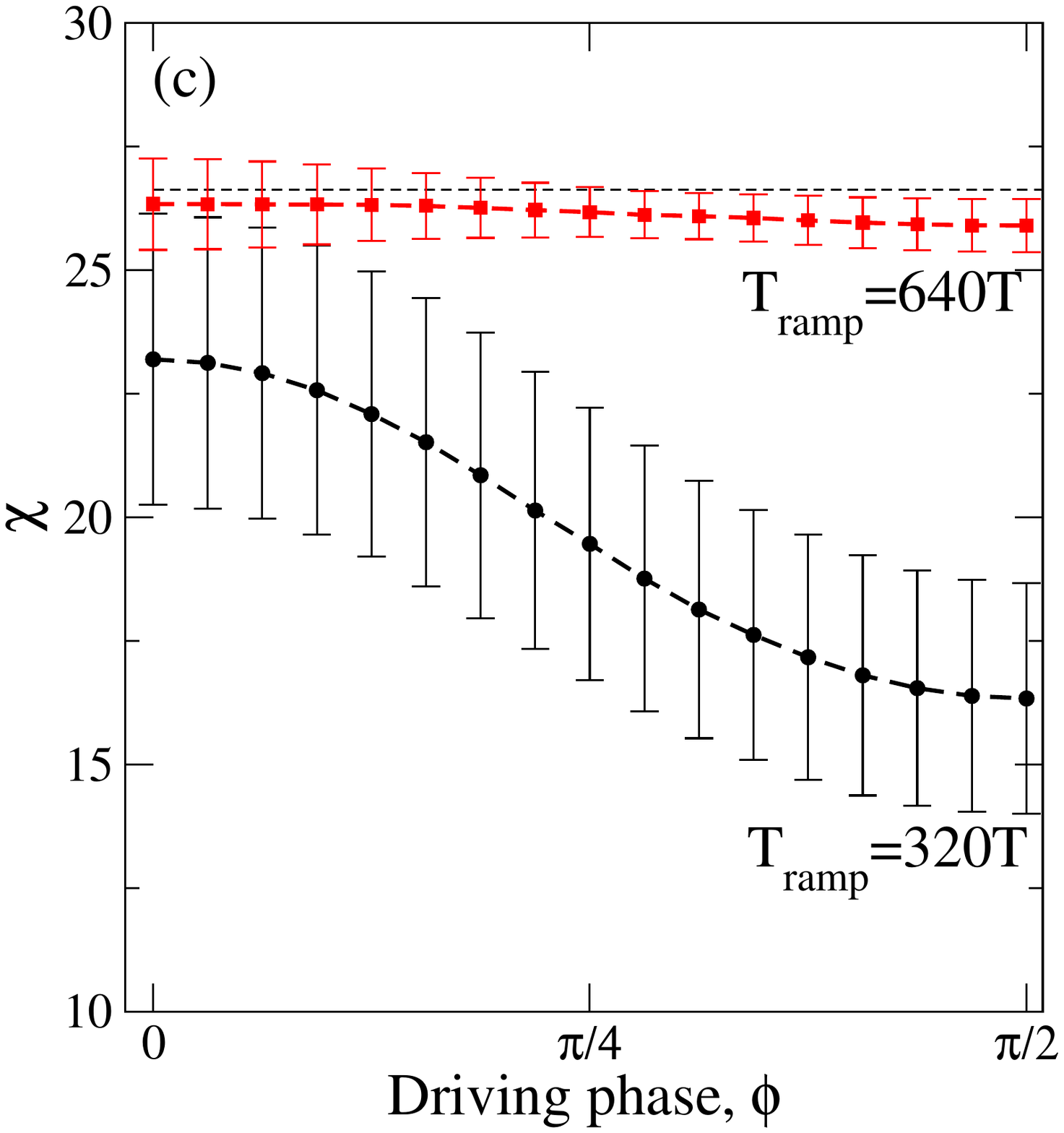}}
\caption{Fidelity of the ramp protocol in preparing the
cat state at $\kappa = 0.8$, as measured by $\chi$.
(a) For a ramp-time of $T_{\mathrm{ramp}}=320 T$
the fidelity of the final state depends strongly on the phase of the driving.
Cosinusoidal driving (black solid line) shows the best performance,
and sinusoidal driving (blue dot-dashed line) the worst. 
An intermediate phase, $f(t)=\cos(\omega t - \pi/4)$ (red dashed line)
lies between these results. The value of $\chi$ in the true ground
state of the system is shown by the horizontal dashed line. 
(b) When the ramp-time is increased to to $T_{\mathrm{ramp}}=640 T$
the driving phase barely affects the result. 
(c) Dependence of $\chi$ on the phase of the driving for the
two ramp-times. Error bars indicate the amplitude of the oscillations
of the final state.
Physical parameters: $U=1$, $\omega=50$.}
\label{ramp}       
\end{figure}

\section{Adiabatic state preparation}
\label{sec:6}

We now consider the feasibility of preparing the cat 
state in experiment, using the technique of adiabatic manipulation.
We initialise the system in the Mott state, with one particle
occupying each site of the lattice, and then gradually increase the value of 
$\kappa$ from zero to $\kappa = 0.8$. If this increase is done
sufficiently slowly, so that the system remains in its ground-state at
all times, the Mott-state will be adiabatically transformed into the
exotic superfluid state. In particular we consider the protocol in
which the amplitude of $\kappa$ is ramped up linearly over a long
time-interval $T_{\mathrm{ramp}}$, which may be many hundred of
driving periods long, and then held at a constant value  for a period 
we term the ``hold-time''.
To quantify the quality of the state preparation, we measure the quantity
$\chi = \langle n_{\pi/2} n_{\pi/2} \rangle$, that is, the
$\left( \pi/2, \pi/2 \right)$ component of the two-particle
reduced density matrix. This quantity takes
a high value in the cat state, and so acts as a good figure of merit \cite{kinetic_PRR}
to identify a state's ``cattiness''.

In Fig. \ref{ramp}a we show the result for sinusoidal driving,
$f(t) = \cos \left( \omega t + \varphi \right)$, for three different
values of the driving phase $\varphi$, using a ramp-time of
$T_{\mathrm{ramp}} = 320 T$. When the driving is cosinusoidal ($\varphi = 0$),
the value of $\chi$ initially rises smoothly, but then begins to oscillate
towards the end of the ramp, and these oscillations continue during the
hold-time. This arises from the form of the quasienergy spectrum of the system.
When the system is in the Mott state, the spectrum is gapped and the 
ground-state is well-separated from the higher energy states. Consequently
the adiabatic condition is easily satisfied and the system
smoothly tracks the instantaneous ground-state as $\kappa$ increases. As the
system passes through the phase transition to the superfluid state, however,
this gap closes, and unless the ramp speed is extremely slow some 
proportion of the state will be excited out of the instantaneous ground state, 
producing the oscillatory
behaviour. Changing the driving-phase to $\varphi=\pi/4$ produces a qualitatively
similar behaviour, but the maximum value of $\chi$ is smaller, indicating
that the final state has been prepared with less fidelity. For a
sinusoidal driving ($\varphi=\pi/2$) the fidelity is reduced even further.

The $\varphi$-dependence of these results may appear surprising, since in 
Section \ref{sec:4}
it was shown that the driving-phase had no effect on the system.
However, that lack of $\varphi$-dependence only applies strictly to the 
ramping procedure in the adiabatic limit, $T_{\mathrm{ramp}} \rightarrow
\infty$. In Fig. \ref{ramp}b we show the effect of doubling the ramp-time to
$T_{\mathrm{ramp}} = 640T$. With this slower ramp, the procedure is more
adiabatic, and indeed the $\varphi$-dependence of the results is 
considerably reduced. As a consequence the maximum value of $\chi$ following
the ramp is much 
closer to the value for the $\kappa=0.8$ superfluid ground state,
and the oscillations in $\chi$ are much reduced. We quantify this further
in Fig. \ref{ramp}{c}, showing the $\varphi$ dependence of $\chi$ for the
slow and the fast ramp-times. Clearly as $T_{\mathrm{ramp}}$ increases,
the $\varphi$ dependence of the results decreases, and the fidelity of the process
improves. From this graph it is also clear that if an experiment is limited 
to low or moderate values of the ramp-time, the best choice of the
driving will be cosenoidal.

\section{Conclusions}
\label{sec:7}
We have investigated the sensitivity of a many-boson system to the details of the kinetic driving previously explored in \cite{kinetic_NJP,kinetic_PRR}, where the hopping energy oscillates periodically in time with zero average.
In particular we have studied the sensitivity of the cat-like structure of the ground state to the shape of the signal profile, phase-shifts of the driving, and the presence of an external magnetic flux. 
We have found that the system is very sensitive to the symmetry of the time signal. The sawtooth profile exemplifies the case where time-reversal symmetry is absent, and the cat-like properties of the ground state are considerably diminished. The fact that the interaction matrix elements between plane waves cannot vanish makes them weakly dependent on the strength of the driving, thus favoring less markedly the collision processes that in \cite{kinetic_PRR} were shown to underlie the ground-state cat structure.

In contrast to the case of the standard driven Bose-Hubbard model, we find that introducing a
phase-shift to the driving signal does not affect the properties of the system at all. Instead of producing hopping phases, as might be expected, the system is completely blind to this form of perturbation. 
Introducing hopping phases by rotating the ring to produce an effective flux,
only has the effect of trivially displacing the momenta at which the quasi-condensates form. In particular, 
the cat structure of the ground state remains intact.

We have also explored the preparation of the cat-state in experiment, by adiabatically ramping the driving from zero. In the deep adiabatic limit, the process becomes insensitive to the phase of the driving, while
a cosenoidal signal gives the optimum fidelity for shorter ramp-times.

In conclusion, the main properties of the kinetically driven superfluid boson system are rather robust against variations in the details of the driving, provided that the signal is time symmetric.
In addition, we have shown that suitably choosing the phase of the driving allows the adiabatic preparation to be substantially shortened.

\section*{Acknowledgments}
This work was supported by Spain's MINECO through
grant FIS2017-84368-P,
and by the UCM through grant FEI-EU-19-12.

\bibliographystyle{aipnum4-1}
\bibliography{prague_bib}

\end{document}